\documentclass[12pt,preprint]{aastex}

\begin{document}
\pagenumbering{arabic}

\title{A STRANGE M\`{E}NAGE \'{A} TROIS}

\author{Sidney van den Bergh}
\affil{Dominion Astrophysical Observatory, Herzberg Institute of Astrophysics, National Research Council of Canada, 5071 West Saanich Road, Victoria, BC, V9E 2E7, Canada}
\email{sidney.vandenbergh@nrc-cnrc.gc.ca}

{\bf  The Magellanic Clouds may have joined our Milky Way system quite recently. The Large Magellanic Cloud turns out to be a remarkably luminous object that is close to the upper luminosity limit of the class of magellanic irregular galaxies.}

 We are all Copernicans now. So we expect to be living in a typical galaxy in a normal neighborhood. The first of these expectations is fulfilled. Our Milky Way system is a relatively normal giant spiral of Hubble type Sbc, or perhaps a barred giant of type SBbc. However, the Galactic neighborhood is unusual and quite different from what might have been expected. True, the Local Group in which we live is a small poor cluster of galaxies like many others in nearby regions of the Universe. However, the nearest neighbors to our home galaxy exhibit two remarkable peculiarities. For most galaxies, like that in Andromeda $^1$ , the nearest neighbors are early-type galaxies of types E or S0, whereas the more distant companions are late-type objects of types Sc or Ir. However, the Milky Way's two closest big companions, the Large Magellanic Cloud (LMC) and the Small Magellanic Cloud (SMC) are irregular galaxies. This anomaly suggests$^2$ that the Magellanic Clouds might not be close satellites of the Galaxy, but objects that formed in the outer reaches of the Local Group that just happen to be passing close to the Milky Way system at the present time. Recent calculations $^3$ suggest that there is a $\sim$72\% probability that the Magellanic Clouds were accreted within the last Gyr, and a $\sim$50\% probability that they were
accreted together.   The second anomaly among the closest
companions to our Galaxy is that the LMC is so extraordinarily luminous for a magellanic irregular galaxy. In nearby regions of the Universe there are only two Ir galaxies (NGC 4214 and NGC 4449) that even come close to rivaling the LMC in luminosity. In other words the Large Magellanic Cloud seems to be close to the upper luminosity limit of $M_B \simeq$-18.5 for irregular galaxies. This is important because there is a fundamental morphological difference between spiral and irregular galaxies: spirals all have nuclei, whereas Magellanic irregulars do not. It should be emphasizes that this upper luminosity limit applies only to magellanic irregulars and not to the peculiar chaotic irregular galaxies that might have been formed during the collisions or mergers of massive ancestral galaxies.

 In 1969 Erik Holmberg $^4$ searched for the satellites of nearby galaxies on the prints of the Palomar Sky Survey.
Surprisingly he found that bright satellite galaxies, like the Magellanic Clouds, are quite rare. This conclusion has recently been strengthened and confirmed by the work of James \& Ivory $^5$ and by Lui et al.$^6$. James \& Ivory used narrow-band imaging of 143 luminous spiral galaxies, comparable to the Milky Way, to search for star forming companions. They concluded that luminous star-forming satellite galaxies are quite rare and that our home galaxy is unusual, both for the luminosity, and the proximity of its two brightest satellites.
A different approach was used by Liu et al. who employed the enormous data-base provided by the Sloan Digital Sky Survey to search host galaxies, with luminosities within $\pm$ 0.2 mag of that of the Milky Way for satellite galaxies with luminosities similar to those of the Magellanic Clouds, that are located within a distance of 150 kpc of of their apparent host galaxy. For 22581 Milky Way-like hosts they find that 81\% have no satellites as bright as the Magellanic Clouds, 11\% have one such satellite, and only 3.5\%  host two such galxies. As Edwin Hubble $^7$ said many years ago `` The fact that the [G]alactic system is a member of a group is a very fortunate accident.'' That the Galaxy should have an irregular companion as luminous as the Large Magellanic Cloud is almost a miracle.

%\begin{references}
\noindent

$^{1}$ Einastro, J. et al. Nature, {\bf 252}, 111-113 (1974)
 
$^{2}$ van den Bergh, S. Astron. J, {\bf 132}, 1571-1574 (2006)

$^{3}$  Busha, M. T. et al. arXiv: 1011,2203v2 (2010)

$^{4}$ Holmberg, E. Ark. Astron. {\bf 5}, 305-343 (1969)

$^{5}$  James, P. A., \& Ivory, C. F. Mon.Not.R.Astron.Soc, 
in press arXiv:1109.2875 (2010)

$^{6}$ Liu, L. et al. arXiv: 1011.2255v2

$^{7}$ Hubble, E. The Realm of the Nebulae, 
New Haven-Yale University press, p.125 (1936)

\end{document}